\documentclass[aps,prl,twocolumn,groupedaddress]{revtex4}

\usepackage{epsfig}
\usepackage{graphicx}
\usepackage{float}
\usepackage{dcolumn}
\usepackage{amsmath}
\input{epsf}

\newcommand{\ket}[1]{\left| {#1} \right\rangle}

\begin{document}

\title{Quantum teleportation with a quantum dot single photon source}


\author{D. Fattal, E. Diamanti}
\author{K. Inoue}
\altaffiliation[Also with: ]{NTT Basic Research Laboratories,
Atsugishi, Kanagawa, Japan.}

\author{Y. Yamamoto}
\altaffiliation[Also with: ]{NTT Basic Research Laboratories,
Atsugishi, Kanagawa, Japan.}


\affiliation{Quantum Entanglement Project, ICORP, JST\\ Ginzton
Laboratory, Stanford University, Stanford CA 94305}


\date{\today}

\begin{abstract}

We report the experimental demonstration of a quantum
teleportation protocol with a semiconductor single photon source.
Two qubits, a target and an ancilla, each defined by a single
photon occupying two optical modes (dual-rail qubit), were
generated independently by the single photon source. Upon
measurement of two modes from different qubits and postselection,
the state of the two remaining modes was found to reproduce the
state of the target qubit. In particular, the coherence between
the target qubit modes was transferred to the output modes to a
large extent. The observed fidelity is 80 \%, a figure which can
be explained quantitatively by the residual distinguishability
between consecutive photons from the source. An improved version
of this teleportation scheme using more ancillas is the building
block of the recent KLM proposal for efficient linear-optics
quantum computation \cite{ref:klm}.

\end{abstract}


\maketitle

Photons are almost ideal carriers of quantum information, since
they have little interaction with their environment, and are easy
to manipulate individually with linear-optics. The main challenge
of optical quantum information processing is the design of
controlled interactions between photons, necessary for the
realization of non-linear quantum gates. Photons do not naturally
"feel" the presence of other photons, unless they propagate in a
medium with high optical non-linearity. The amount of optical
non-linearity required to perform controlled operations between
single photons is however prohibitively large. 

Probabilistic gates can be implemented with linear-optics only
\cite{ref:Ralph, ref:klm, ref:Franson}, but as such, they are not
suitable for scalable quantum computation. In a seminal paper
\cite{ref:Gottesman}, Gottesman and Chuang suggested that quantum
gates could be applied to photonic qubits through a generalization
of quantum teleportation \cite{ref:telep}. In such a scheme, the
information about the gate is contained in the state of ancilla
qubits. The implementation of a certain class of gates can then be
reduced to the problem of preparing the ancilla qubits in some
wisely chosen entangled state. Such a problem can be solved
"off-line" with linear-optics elements only, provided the photons
used are quantum mechanically indistinguishable particles
\cite{ref:Innsbruck}. Following this idea, Knill, Laflamme and
Milburn (KLM) \cite{ref:klm} proposed a scheme for efficient
linear-optics quantum computation (LOQC) based on the
implementation of the controlled-sign gate (C-z gate) through
teleportation. Since the C-z gate acts effectively on only one of
the two modes composing the target qubit, a simplified procedure
can be used where a single optical mode is teleported, instead of
the two modes composing the qubit.

This procedure will be referred to as \textit{single mode
teleportation} to distinguish it from the usual teleportation
scheme. In its basic version using one ancilla qubit (ie two
ancilla modes), this procedure succeeds half of the time. In its
improved version using an arbitrarily high number of ancillas, it
can succeed with a probability arbitrarily
close to one \cite{ref:klm,ref:Franson1}.\\

In this paper, we report an experimental demonstration of the
basic version of the single mode teleportation. We use quantum
mechanically indistinguishable photons from a quantum dot single
photon source \cite{ref:Charlie01}, featuring high suppression of
two-photon pulses. The fidelity of the teleportation depends
critically on the quantum indistinguishability of two photons
emitted independently by the single photon source, a feature that
was experimentally verified in \cite{ref:Charlie02}. A similar
experiment was done in the past using two photons emitted
spontaneously by parametric down conversion (PDC)
\cite{ref:DeMartini}. However, the efficiency of such a process is
intrinsically limited by the presence of two-photon pulses, which
makes it unsuitable when more identical photons are needed, e.g.
to implement the improved teleportation scheme. To date, the
demonstration of the single mode teleportation with a single
photon source remained a capital step to be achieved towards
scalable LOQC.

\begin{figure}[hbtp]
    \epsfig{file=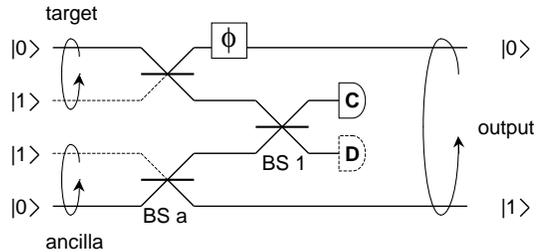, bbllx=130, bblly=70, bburx=330, bbury=460,  angle=-90, width=3in, clip=}
    \caption{Schematic of single mode teleportation. Target and ancilla qubits are each defined by a single
    photon occupying two optical modes. When detector C clicks and D does not, the state of the remaining modes reproduces the state
    of the target. The coherence between modes (1) and (2) of the target was transferred to a coherence between the same mode
    (1)of the target and mode (4) of the ancilla. Preparing the target in an equal superposition state makes it easier to measure the transfer of coherence.}
    \label{fig:scheme}
\end{figure}

The single mode teleportation in its simplest form involves two
qubits, a target and an ancilla, each defined by a single photon
occupying two optical modes (see fig \ref{fig:scheme}). The target
qubit can a priori be in an arbitrary state $\alpha \ket{0}_L +
\beta \ket{1}_L$ where the logical $\ket{0}_L$ and $\ket{1}_L$
states correspond to the physical states $\ket{1}_1 \ket{0}_2$ and
$\ket{0}_1 \ket{1}_2$ respectively in a dual rail representation.
The ancilla qubit is prepared with a beam-splitter (BS a) in the
coherent superposition $\frac{1}{\sqrt{2}}(\ket{0}_L+\ket{1}_L) =
\frac{1}{\sqrt{2}}(\ket{1}_3 \ket{0}_4 + \ket{0}_3 \ket{1}_4)$.
One rail of the target (mode 2) is mixed with one rail of the
ancilla (mode 3) with a beam-splitter (BS 1), for subsequent
detection in photon counters C and D. For a given realization of
the procedure, if only one photon is detected at detector C, and
none at detector D, then we can infer the resulting state for the
output qubit - composed of mode (1) and (4) :
\[ \psi_C = \alpha \ket{0}_L + \beta  \ket{1}_L = \alpha \ket{1}_1\ket{0}_4 + \beta  \ket{0}_1\ket{1}_4  \] which
is the initial target qubit state. Similarly, if D clicks and C
does not, then the output state is inferred to be:
\[ \psi_D = \alpha \ket{0}_L - \beta  \ket{1}_L = \alpha \ket{1}_1\ket{0}_4 - \beta  \ket{0}_1\ket{1}_4 \]
which again is the target state up to an additional phase shift of
$\pi$, which can be actively corrected for \cite{ref:DeMartini2}.
We did not implement this active feedforward here. Half of the
time, either zero or two photons are present at counters C or D,
and the teleportation procedure fails. It is interesting and
somewhat enlightening to describe the same procedure in the
framework of \textit{single rail logic}. In this framework, each
optical mode supports a whole qubit, encoded in the presence or
absence of a photon, and the single mode teleportation can be
viewed as entanglement swapping. Indeed, for the particular values
$\alpha = \beta = \frac{1}{\sqrt{2}}$ modes 1 and 2 find
themselves initially in the Bell state $\ket{\psi^+}_{12}$, while
modes 3 and 4 are in a similar state $\ket{\psi^+}_{34}$. A
partial Bell measurement takes place using BS 1 and counters C/D,
which if it succeeds leaves the system in the entangled state
$\ket{\psi^+}_{14}$, so that entanglement swapping occurs. In the
rest of the paper, we choose to consider the scheme in the dual
rail picture, since it is a more robust, hence realistic way of
storing quantum information (at the expense of using two modes per
qubit).

The success of the teleportation depends mainly on the transfer of
coherence between the two modes of the target qubit. If the target
qubit is initially in state $\ket{0}_L = \ket{1}_1\ket{0}2$, then
the ancilla photon cannot end up in mode (4) because of the
postselection condition, so that the output state is always
$\ket{1}_1\ket{0}4$ as wanted. The same argument applies when the
target qubit is in state $\ket{1}_L$. However, when the target
qubit is in a coherent superposition of $\ket{0}_L$ and
$\ket{1}_L$, the output state might not retrieve the full initial
coherence. We can test the transfer of coherence by preparing the
target in a maximal superposition state:
\[ \psi_{tar} = \frac{1}{\sqrt{2}}\left({\ket{0}_L+e^{i \phi} \ket{1}_L}\right) \]
where $\phi$ is a phase that we can vary. If the initial coherence
of the target qubit is not transferred to the output qubit, a
change in $\phi$ will not induce any measurable change in the
output qubit. However, if changing $\phi$ induces some measurable
change in the output qubit, then we can prove that the initial
coherence was indeed transferred, at least to some extent.\\

\begin{figure}[hbtp]
    \epsfig{file=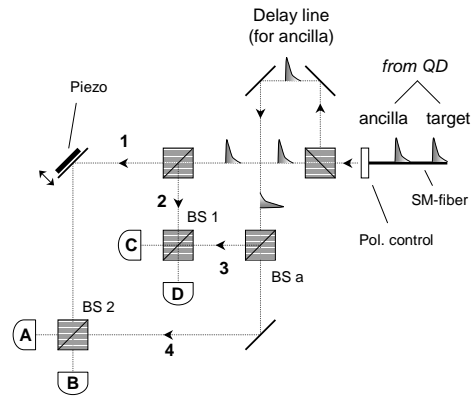, bbllx=100, bblly=140, bburx=470, bbury=600,  angle=-90, width=3in, clip=}
    \caption{Experimental setup. All the beam-splitters (BS) shown are 50-50 non-polarizing BS. The teleportation procedure works when the ancilla photon is delayed,
    but the target is not. After preparation, the target photon occupies modes 1 and 2, and the ancilla occupies
    modes 3 and 4. Modes 2 and 3 are mixed at BS 1 and subsequently measured by detectors C and D, this step being the
    heart of the teleportation. When C clicks and D does not, a single photon occupies modes 1 and 4 which constitute the
    output qubit. The relative phase between modes 1 and 4 in the
    output state is measured by mixing those modes at BS 2 and
    recording single counts at detector A or B. Note that since
    an event is recorded only if A \textit{and} C or B \textit{and}
    C clicked, the condition that D did not click is automatically
    fulfilled.    }
    \label{fig:setup}
\end{figure}

The experimental setup is shown in fig. \ref{fig:setup}. Two
photons emitted consecutively by the single quantum dot photon
source \cite{ref:Charlie01,ref:Charlie02} are captured in a single
mode fiber. In the dual rail representation, we refer to the first
photon as the ancilla, and to the second photon as the target (see
fig \ref{fig:scheme}). The ancilla qubit, initially in state
$\ket{0}_L$, is delayed in free space to match the target qubit
temporally at BS 1. The delay must be adjusted to within a
fraction of the photons temporal width ($\sim$ 200 ps or 6 cm in
space). Note that the mode matching is significantly easier here
than in similar experiments using photons from PDC, where the
optical path length have to be adjusted with a tolerance of only a
few microns \cite{ref:DeMartini}.

The ancilla is prepared in the superposition state
$\psi_{anc}=\frac{1}{\sqrt{2}}\left({\ket{0}_L+\ket{1}_L}\right)$
with a beam-splitter 'BS a'. The target qubit is prepared in a
similar superposition state, with in addition a variable phase
between the two modes, so that $\psi_{tar} =
\frac{1}{\sqrt{2}}\left({\ket{0}_L+e^{i \phi} \ket{1}_L}\right)$.
The phase shift is applied by changing the path length on mode (1)
with a piezo-actuated miror. The "partial Bell measurement"
responsible for the teleportation is done at BS 1 by mixing the
optical modes (2) of the target qubit and (3) of the ancilla
qubit, with subsequent detection in counter C. A Mach-Zehnder type
setup is used to measure the coherence between the two modes (1)
and (4) of the output qubit. It is composed of a 50-50
beam-splitter BS 2 mixing modes (1) and (4), with subsequent
detection in counters A and/or B. Modulating the phase $\phi$ of
the target qubit should result in the modulation of the count rate
in detector A and B (conditioned on a click at detector C), with a
contrast related to the degree of coherence between the output
modes (1) and (4).

\begin{figure}[hbtp]
    \epsfig{file=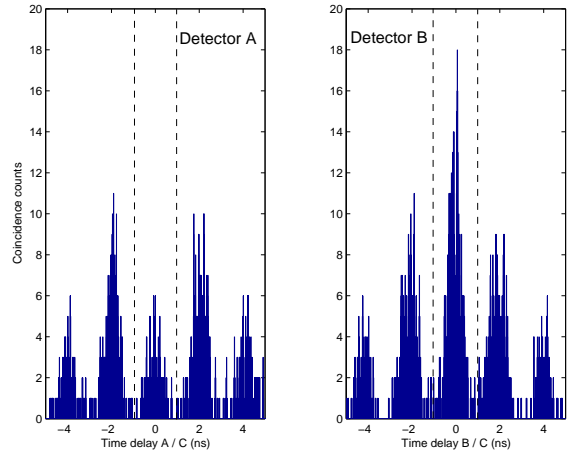, width=3in}
    \caption{typical correlation histograms taken simultaneously between
    detectors A/C and B/C. The central region indicated by the dashed lines
    correspond to the postselected events, when target and ancilla photons had
    such a timing that it is impossible to distinguish between them based on
    the time of detection. As the phase $\phi$ varies, so does the relative size
    of the central peaks for detector A and B. The sum of count rates for the central
    peaks of detector A and B was 800 /s, independently of $\phi$ as shown in fig \ref{fig:result}.}
    \label{fig:example}
\end{figure}

\begin{figure}[hbtp]
    \epsfig{file=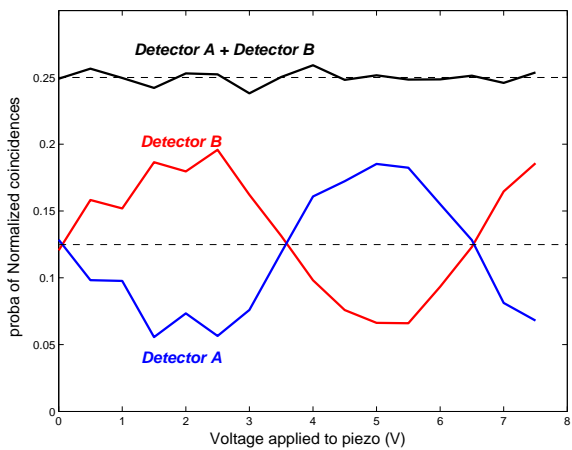, width=3in}
    \caption{Verification of single mode teleportation.
    Coincidence counts between detector A/C and B/C are plotted
    for different voltages applied to the piezo transducer, i.e.
    for different phase $\phi$ of the target qubit. The observed modulation of
    the counts implies that the initial coherence contained in the
    target qubit was transferred to a large extent to the output
    qubit. The reduced contrast ($\sim 60\%$) is principally due to imperfect indistinguishability between
    the target and ancilla photons.}
    \label{fig:result}
\end{figure}

Coincidences between counters A-C and B-C were simultaneously
recorded, by using a start-stop configuration (each electronic
"start" pulse generated by counter C was doubled for this
purpose). This detection method naturally post-selects events
where one photon went through BS 1, and the other went through BS
2, as required by the teleportation scheme. Since no more than one
photon is emitted by the single photon source, no photon can reach
detector D. Typical correlation histograms are shown in fig
\ref{fig:example}. The integration time was 2 min, short enough to
keep the relative optical path length between different arms (1-4)
of the interferometer stable. The whole setup was made compact for
that purpose, and stability over time periods as long as 10 min
was observed. A second post-selection was made, depending on the
timing between target and ancilla photons, which is adequate only
one time out of four - the ancilla taking the long path and the
target the short path. The resulting coincidence counts were
recorded for different phases $\phi$ of the target qubit. The
result of the experiment is shown in fig. \ref{fig:result}. The
number of counts recorded in the post-selected window (-1 ns
$<\tau<$ 1 ns) was normalized by the total number of counts
recorded in detectors A and B in the broader window -5 ns $<\tau<$
5 ns, corresponding to all events where one photon went through BS
1 and the other through BS 2 (but only one quarter of the time
with right timing). Complementary oscillations are clearly
observed at counter A and at counter B, indicating that the
initial coherence was indeed transferred to the output qubit. In
other words, mode (2) of the target qubit
was "replaced" by mode (4) of the ancilla without a major loss of coherence.\\

If the initial coherence was fully conserved during the single
mode transfer, the state of the output qubit would truly be
$\alpha \ket{0}_L + \beta \, e^{i\phi} \ket{1}_L$, and the single
count rate at detector A (resp. B) would be proportional to
$cos^2(\frac{\phi}{2})$ (resp. $sin^2(\frac{\phi}{2})$), giving a
perfect contrast as the target phase $\phi$ is varied. More
realistically, part of the coherence can be lost in the transfer,
resulting in a degradation of the contrast. Such a degradation is
visible on fig. \ref{fig:result}. It arises mainly due to a
residual distinguishability between ancilla and target photons.
Slight misalignments and imperfections in the optics also result
in an imperfect mode matching at BS 1 and BS 2, reducing the
contrast further. Finally, the residual presence of two-photon
among pulses can reduce the contrast even more, although this
effect is negligible here. The overlap V = $\int \psi_{tar} \,
\psi_{anc}$ between target and ancilla wave-packets
\cite{ref:Charlie02}, the two-photon pulses suppression factor
$g^{(2)}$ \cite{ref:Charlie01}, as well as the non-ideal mode
matching at BS 1 and BS 2 - characterized by the first-order
interference visibilities $V_1$, $V_2$ - were all measured
independently. The results are $V \sim 0.75$ (measured with the
setup described in \cite{ref:Charlie02}), $g^{(2)}(0) \sim 2\%$,
$V_1 \sim 0.92$ and $V_2 \sim 0.91$. The contrast $C$ in counts at
detector A or B when we vary the phase $\phi$ should be:
\[ C = \frac{V \cdot V_1 \cdot V_2}{1 + g^{(2)}/2} \sim 0.62 \]
This predicted value compares well with the experimental value of
$C_{exp} \sim 0.60$.

The fidelity of teleportation is $F = \frac{1 + C}{2} \sim 0.8$.
This high value is still not enough to meet the requirements of
efficient LOQC \cite{ref:klm}. In particular, the quantum
indistinguishability of the photons must be increased further to
meet these requirements. In our single photon source, a dephasing
mechanism acting on a time scale of a few nanoseconds
\cite{ref:pss_conf} is responsible for the loss of
indistinguishability. Using the Purcell effect
\cite{ref:YYpaper2}, one can reduce the quantum dot radiative
lifetime well below this dephasing time. However, current jitter
in the photon emission time will eventually prevent any further
reduction of the quantum dot lifetime. Time jitter happens as a
consequence of the incoherent character of our method to excite
the quantum dot \cite{ref:Charlie01}. It is currently of order 10
ps. Time jitter can be completely suppressed using a coherent
excitation technique (see e.g. \cite{ref:Kuhn} for such a scheme
with single atoms). It therefore seems important to develop such
techniques on single quantum dots.

 Using more ancillas in a scheme first proposed in \cite{ref:klm} and
significantly improved in \cite{ref:Franson1}, the single mode
teleportation can be made nearly deterministic. This would allow
the replacement of deterministic non-linear gates necessary for
scalable quantum computation with probabilistic ones, recently
demonstrated experimentally with linear optics \cite{ref:Franson}.
This generalized teleportation procedure requires more
indistinguishable ancilla photons, produced no more than one at a
time, a feature absent in \cite{ref:DeMartini} but present in our
implementation of the teleportation. We also point out that the
generalized scheme requires the discrimination of different photon
numbers. Progress in this direction were reported in
\cite{ref:VLPC}, in which photon numbers up to six could be
discriminated. This would in principle allow the implementation of
a linear-optics C-z gate with a success probability of
$\left( \frac{6}{7} \right)^2 \sim 0.73$ \cite{ref:Franson1}.\\

In conclusion, we have demonstrated the basic version of the
single mode teleportation procedure described in the KLM scheme
with independent single photons and linear-optics. LOQC has
emerged in recent years as an appealing alternative to previous
quantum computation schemes, and to date there had been no
experimental proof of principle except for those based on
parametric down conversion, a technique that sets limits to the
scalability of the system. Our experiment suggests that it is
possible to build an efficient QIP unit using single photon
sources and linear-optics, provided the photons generated are
indistinguishable.


\end{document}